\documentclass[runningheads]{llncs}
\usepackage{graphicx}
\usepackage{bbding}
\usepackage{amsmath}
\usepackage{amssymb}
\begin{document}
\title{Annotation-Free Cardiac Vessel Segmentation via Knowledge Transfer from Retinal Images}
\titlerunning{Annotation-Free Cardiac Vessel Segmentation}
%
\author{Fei Yu\inst{1} \and 
	Jie Zhao\inst{2} \and 
	Yanjun Gong\inst{3} \and 
	Zhi Wang\inst{3} \and 
	Yuxi Li\inst{3} \and 
	Fan Yang\inst{3} \and 
	Bin Dong\inst{4,1,2} \and 
	Quanzheng Li\inst{5,2} \and 
	Li Zhang\inst{1,2(}\Envelope\inst{)}}
%
\authorrunning{F. Yu et al.}
%
\institute{Center for Data Science, Peking University, Beijing 100871, China; \\
	\email{zhangli\_pku@pku.edu.cn} \and
	Center for Data Science in Health and Medicine, Peking University,\\ Beijing 100871, China; \and
	Department of Cardiology, Peking University First Hospital, Beijing 100034, China; \and
	Beijing International Center for Mathematical Research (BICMR), Peking University, Beijing 100871, China; \and
	MGH/BWH Center for Clinical Data Science, Boston, MA 02115, USA. }
\maketitle              
\begin{abstract}
Segmenting coronary arteries is challenging, as classic unsupervised methods fail to produce satisfactory results and modern supervised learning (deep learning) requires manual annotation which is often time-consuming and can some time be infeasible. To solve this problem, we propose a knowledge transfer based shape-consistent generative adversarial network (SC-GAN), which is an annotation-free approach that uses the knowledge from publicly available annotated fundus dataset to segment coronary arteries. The proposed network is trained in an end-to-end fashion, generating and segmenting synthetic images that maintain the background of coronary angiography and preserve the vascular structures of retinal vessels and coronary arteries. We train and evaluate the proposed model on a dataset of 1092 digital subtraction angiography images, and experiments demonstrate the supreme accuracy of the proposed method on coronary arteries segmentation.

\keywords{Coronary Artery Segmentation \and Knowledge Transfer \and Generative Adversarial Network \and Deep Learning.}
\end{abstract}
\section{Introduction}
Quantitative measurement of coronary arteries in medical images is important for the diagnosis, prevention and therapeutic evaluation of related diseases including hypertension, myocardial infarction, and coronary atherosclerotic disease. In the diagnosis of coronary diseases, digital subtraction angiography (DSA) has been widely used and considered the "gold standard". To quantitatively segment blood vessels in DSA, researchers have developed automated methods including region growing, level sets, and Hessian analysis \cite{adams1994seeded,osher1988fronts}. However, due to the complexity of the vascular morphology, these unsupervised methods are difficult to obtain a clinically satisfactory segmentation of coronary arteries. On the other hand, supervised learning such as deep neural networks (DNNs) can produce better segmentation results but relies heavily on pixel-level image annotation, which is often expensive, time-consuming, and even impossible to access especially for coronary artery segmentation.

To solve the problem of the lack of DSA vessel annotation, we aim to apply a strategy to transfer the knowledge of retinal vessel segmentation to the coronary artery segmentation. The researchers have established and validated a number of public retinal vessel segmentation datasets, including DRIVE\cite{staal2004ridge}, STARE\cite{hoover2003locating}, and RITE\cite{hu2013automated}. Due to the significant differences between their anatomical regions, traditional transfer methods are not suitable. Therefore, we present a novel knowledge transfer based adversarial model containing three parts called generator, discriminator, and segmentor. Training of the model contains three major steps: 1) Frangi vessel analysis \cite{frangi1998multiscale} is used to segment the coronary artery in the DSA images roughly. 2) The adversarial training between generator and discriminator allows the model to fuse the fundus image and the DSA image. 3) A synthetic label is then created by computing the union of rough coronary artery segmentation and retinal vessel annotation. Moreover, a shape-consistent scheme is used to ensure the shape consistency of synthetic images and synthetic annotations. The fused image and corresponding synthetic label are used to train the segmentor. The supreme accuracy demonstrates the effectiveness of our methods, which improves the accuracy of coronary artery segmentation by using the knowledge of fundus segmentation without additional manual annotation of the DSA images. The ideas of knowledge transfer and data fusion in this paper have many other application scenarios, including cell segmentation, neural segmentation, and airway segmentation. As long as the object structures in the two datasets are similar, we can use the knowledge from the annotated dataset to guide the analysis of the unannotated one.

Our work relates closely to the recent rise of knowledge transfer techniques. In the field of natural image analysis, Domain-Adversarial Neural Network (DANN) transfers the feature distribution to solve the domain-shift problem \cite{ganin2016domain}. Cycle-GAN introduces a cycle consistency loss and achieves unpaired image-to-image translation \cite{zhu2017unpaired}. AdaptSegNet adopts adversarial learning in the output space and receives favorably accuracy and visual quality \cite{tsai2018learning}. In the field of medical image analysis, many studies have been dedicated to exploring cross-modality translation with GAN \cite{nie2017medical}. Using synthetic data to overcome insufficient labeled data is also an active research area. For example, using synthetic data as augmented training data can help lesion segmentation \cite{kamnitsas2017unsupervised} and cardiovascular volumes segmentation \cite{zhang2018translating}. These methods show that knowledge transfer is effective for the same anatomical region, and our work contributes by accomplishing knowledge transfer between two different anatomical regions.

\section{Methods}
In this section, we show two models based on knowledge transfer. The first one is the GAN model with a constraint of shape consistency (SC-GAN) proposed in this work. Then, to verify the necessity of fusion, we show another simpler model. It adopts from Mixup\cite{zhang2017mixup}, which is to train the U-Net model by computing an average of the fundus image and the DSA image (Add U-Net).

\subsection{SC-GAN}
\begin{figure}
	\centering
	\includegraphics[width =\textwidth]{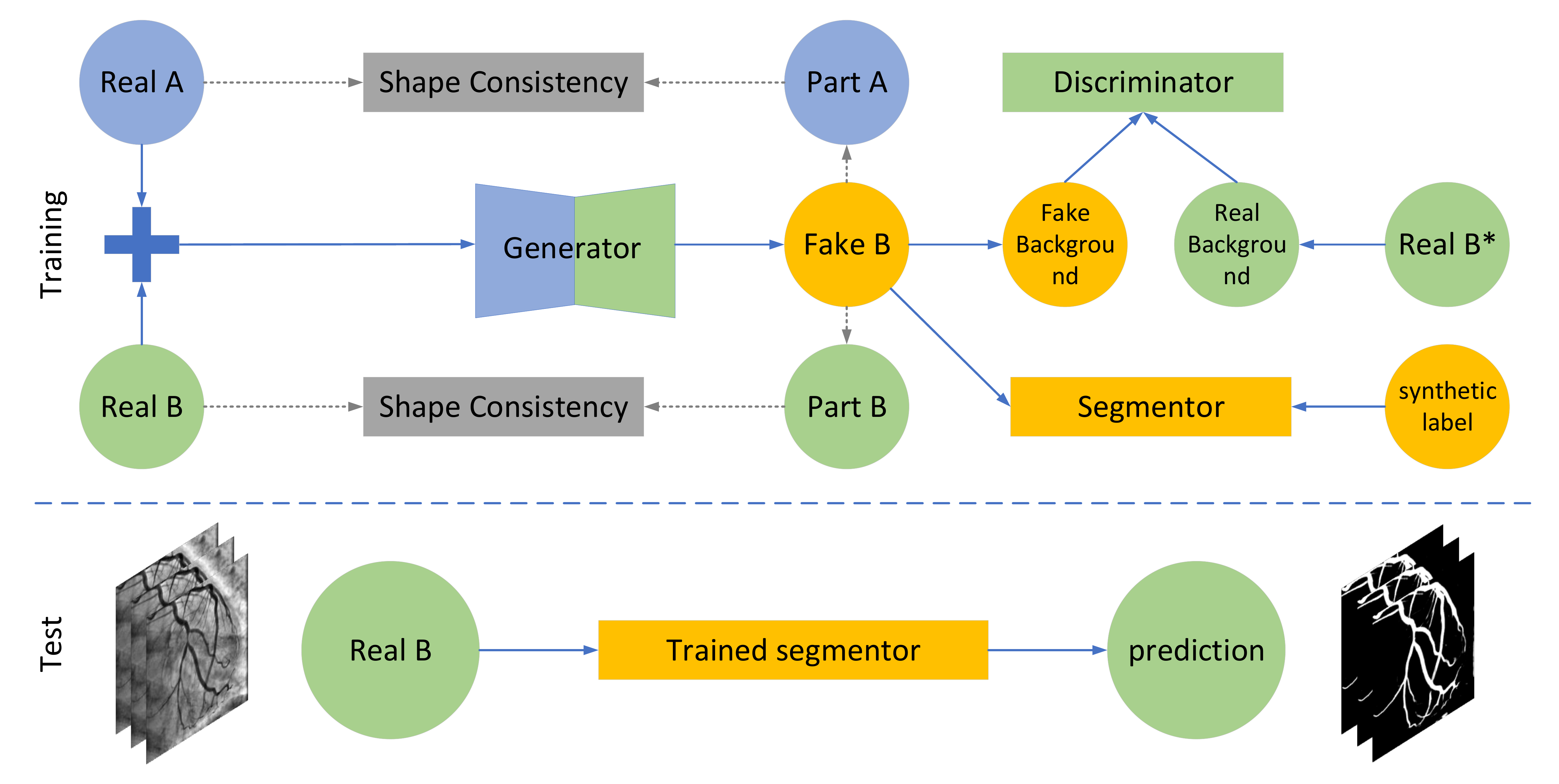}
	\caption{The illustration of the proposed SC-GAN, an end-to-end approach for coronary artery segmentation requiring no new manual annotations.}
	\label{fig:network}
\end{figure}
Figure \ref{fig:network} shows the training and test processes of the proposed model. During the training process, the generator, discriminator, and segmentor are trained simultaneously. In the test process, only the trained segmentor is needed to segment coronary arteries, which requires much less memory and inference time than training. In the previously reported knowledge transfer between different modalities within the same anatomy, the foreground and the background can be reasonably registered. In our task, however, the foreground and background of the images are completely mismatched. So we have designed a shape-consistent scheme that allows the generator and discriminator to complete the knowledge transfer of foreground and background respectively. In the following subsections, we will discuss the model's architectures and the objective functions in detail.

\paragraph{Generator:} As shown in Figure \ref{fig:network}, the generator uses U-Net as its network backbone. The input of the generator is an average of the fundus image (Real A) and the DSA image (Real B). The output is a synthetic image (Fake B) with the same dimensions as the input. To ensure that the Fake B has both retinal vessels and coronary arteries, we extract the two vascular regions (Part A and Part B) in Fake B using manual annotation of retinal vessels and Frangi segmentation of DSA images. A shape-consistent loss ($l1$ loss) is then used to regularize the content of the vessel regions in Fake B to be consistent with the corresponding regions in the original images,
\begin{equation}
\mathcal{L}_{shape_A}(G)=\mathbb{E}_{A \sim p_{data}(A), B \sim p_{data}(B)}[||label_A*(G(A+B)-A)||_1]
\end{equation}
\begin{equation}
\mathcal{L}_{shape_B}(G)=\mathbb{E}_{A \sim p_{data}(A), B \sim p_{data}(B)}[||label_B*(G(A+B)-B)||_1]
\end{equation}
where $A$ represents the fundus image, $B$ represents the DSA image, and $label_A$ and $label_B$ are the retinal vessel annotation and the Frangi segmentation results, respectively.

\paragraph{Discriminator:} We expect that the background of Fake B is sufficiently similar to the background of a DSA image, so we first use the annotations and Frangi analysis results to mask out the vascular regions and extract the background in the generated and real DSA images,
\begin{equation}
Fake_{bg}=\neg(label_A \cup label_B)*G(A+B)
\end{equation}
\begin{equation}
Real_{bg}=\neg(label_A \cup label_B)*B^*
\end{equation}
Where $B^*$ (Real $B^*$ in Figure \ref{fig:network}) represents a randomly chosen DSA image before the injection of the contrast medium, which guarantees a vessel-free background. For the structure of the discriminator, we adopt PatchGAN \cite{zhu2017unpaired}.  The adversarial loss between the generator and the discriminator can be expressed as,
\begin{equation}
\begin{split}
\mathcal{L}_{GAN}(G,D)&=\mathbb{E}_{B^* \sim p_{data}(B^*)}[logD(Real_{bg})]\\
&+\mathbb{E}_{A \sim p_{data}(A), B \sim p_{data}(B)}[log(1-D(Fake_{bg}))]
\end{split}
\end{equation}

\paragraph{Segmentor:} The main structure of the segmentor is also a U-Net. We use MultiLabelSoftMarginLoss \cite{lapin2018analysis} as the objective function of the segmentor:
\begin{equation}
\mathcal{L}_{seg}(S)=-(y(i)log[\frac{exp(\hat{y}(i))}{1+exp(\hat{y}(i))}]+(1-y(i))log[\frac{1}{1+exp(\hat{y}(i))}])
\end{equation}
where $\hat{y}$ is the prediction and $y$ the synthetic label.

Therefore, the final objective function of our proposed model is,
\begin{equation}
\mathcal{L}(G,D,S)= \mathcal{L}_{GAN}(G,D) + \mathcal{L}_{seg}(S) + \lambda\mathcal{L}_{shape_A}(G) + \mu \mathcal{L}_{shape_B}(G) 
\end{equation}
where $\lambda$ and $\mu$ control the relative importance of the objectives. During training, we set $\lambda=100,\mu=50$.

The model uses instance normalization \cite{ulyanov2016instance} instead of batch normalization \cite{ioffe2015batch}, the generator uses ReLU and the discriminator uses LeakyReLU as activations.

\subsection{Add U-Net}
\begin{figure}
	\centering
	\includegraphics[width =0.9\textwidth]{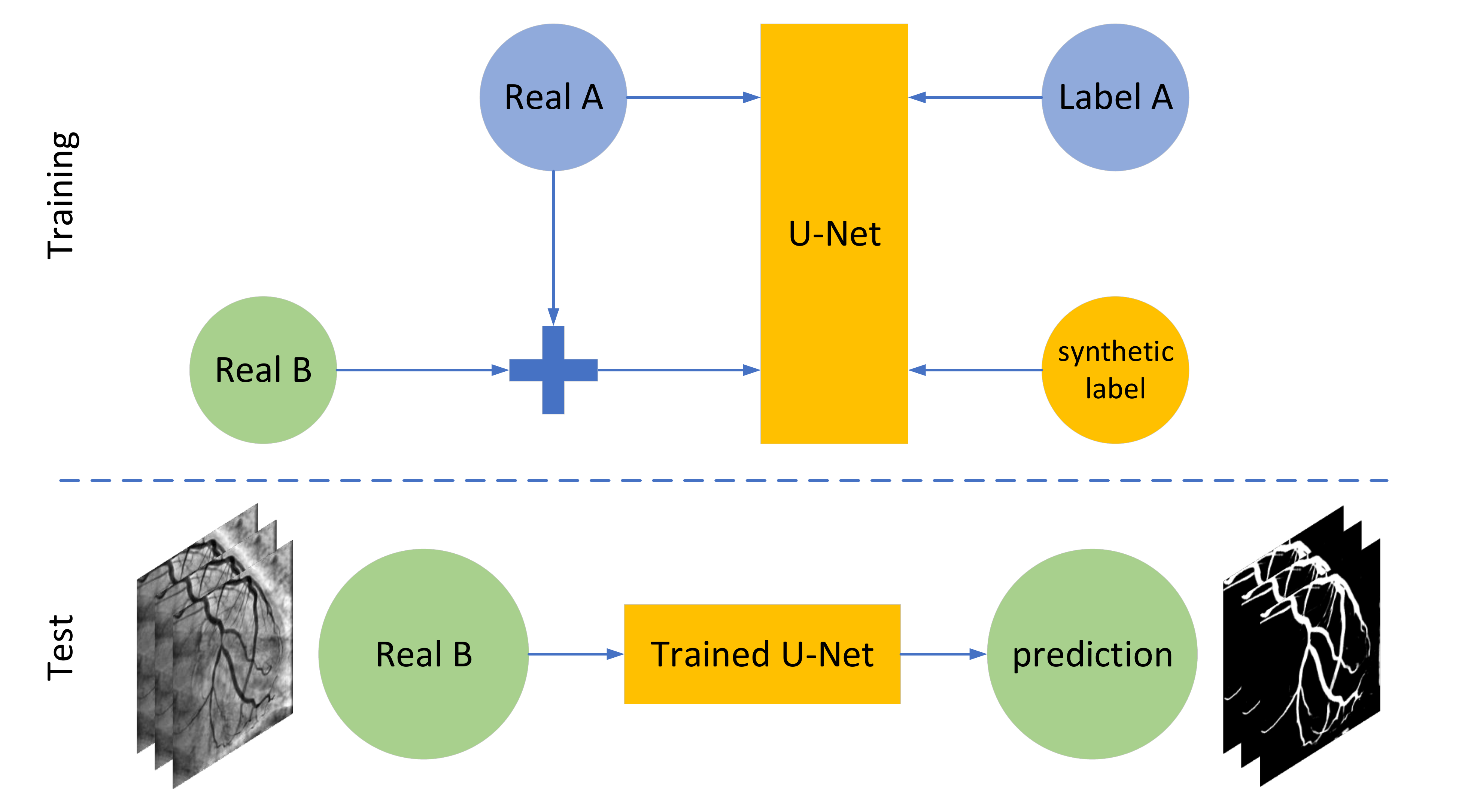}
	\caption{The illustration of Add U-Net, where the input is an average of fundus photography and DSA image.}
	\label{fig:addimage}
\end{figure}

Figure \ref{fig:addimage} shows the overall structure of Add U-Net. Referring to Mixup, the model takes the average of the fundus image (Real A) and the DSA image (Real B). The manual annotation of retinal vessels and the Frangi analysis results of the DSA image are combined to obtain the label of the added image. We then train a U-Net with such  added images and annotations. An independent DSA dataset is used to evaluate the trained U-Net.

\section{Experiments}
In this section, to evaluate the effectiveness of our proposed SC-GAN, we compare the segmentation of four methods: \textbf{1) Frangi Algorithm} Multi-scale Frangi vessel analysis is used to segment coronary arteries. \textbf{2) Classic U-Net}. A U-Net model is trained using Frangi Algorithm results as the learning targets. \textbf{3) Add U-Net}. A U-Net model is trained using an average of fundus photographies and DSA images. \textbf{4) SC-GAN} The proposed shape-consistent GAN. We also compare the synthetic results of SC-GAN and Cycle-GAN \cite{zhu2017unpaired}. The data and experiment details are presented below.
\paragraph{Data:}
We use the DRIVE \cite{staal2004ridge} dataset as the source domain of knowledge transfer. The DRIVE dataset includes 40 fundus images with manually annotated retinal vessels. We also collect 1092 coronary angiographies (DSA) with no annotations as the target domain of knowledge transfer. Several preprocessing approaches are performed on the fundus images, including color to grayscale transform, median-filtering and contrast-limited adaptive histogram equalization \cite{zuiderveld1994contrast}. Finally, we resize all images into the same size of 512$\times$512 and randomly choose 256$\times$256 patches as inputs of the models.
\paragraph{Experiment details:}
In all experiments, 50\% of DSA images are randomly selected as training set, 20\% are validation set and 30\% are test set. Meanwhile, we use the Adam solver \cite{kingma2014adam} with a learning rate of 2e-4. After training for 50 epochs, we decrease the learning rate linearly for 50 epochs till 0.

\section{Results}
\begin{figure}
	\centering
	\includegraphics[width=0.7\textwidth]{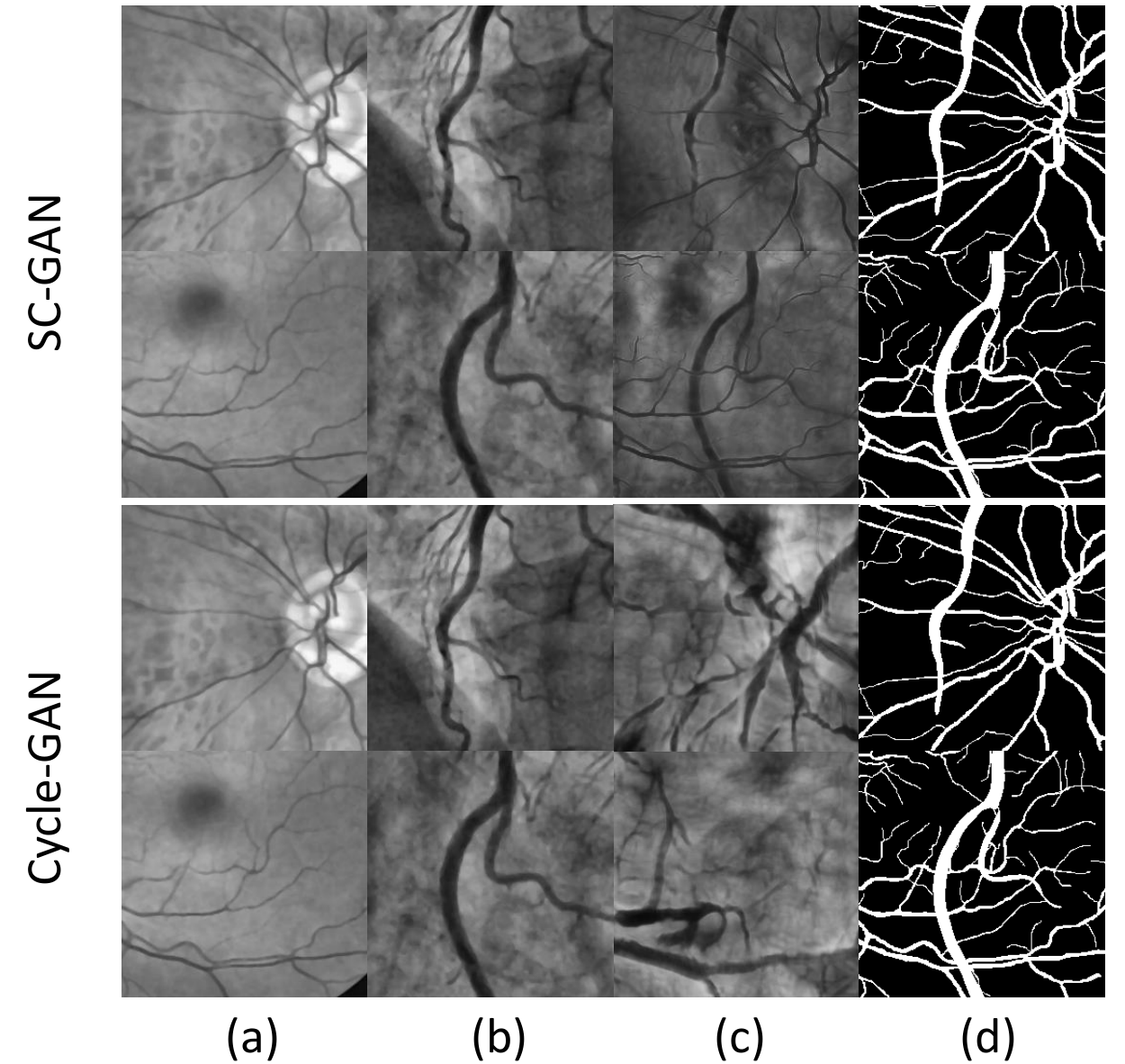}
	\caption{Comparison of SC-GAN and Cycle-GAN.(a) Fundus patches, (b) DSA patches, (c) synthetic images, (d) synthetic labels.}
	\label{fig:translation}
\end{figure}

\begin{figure}
	\centering
	\includegraphics[width =0.8\textwidth]{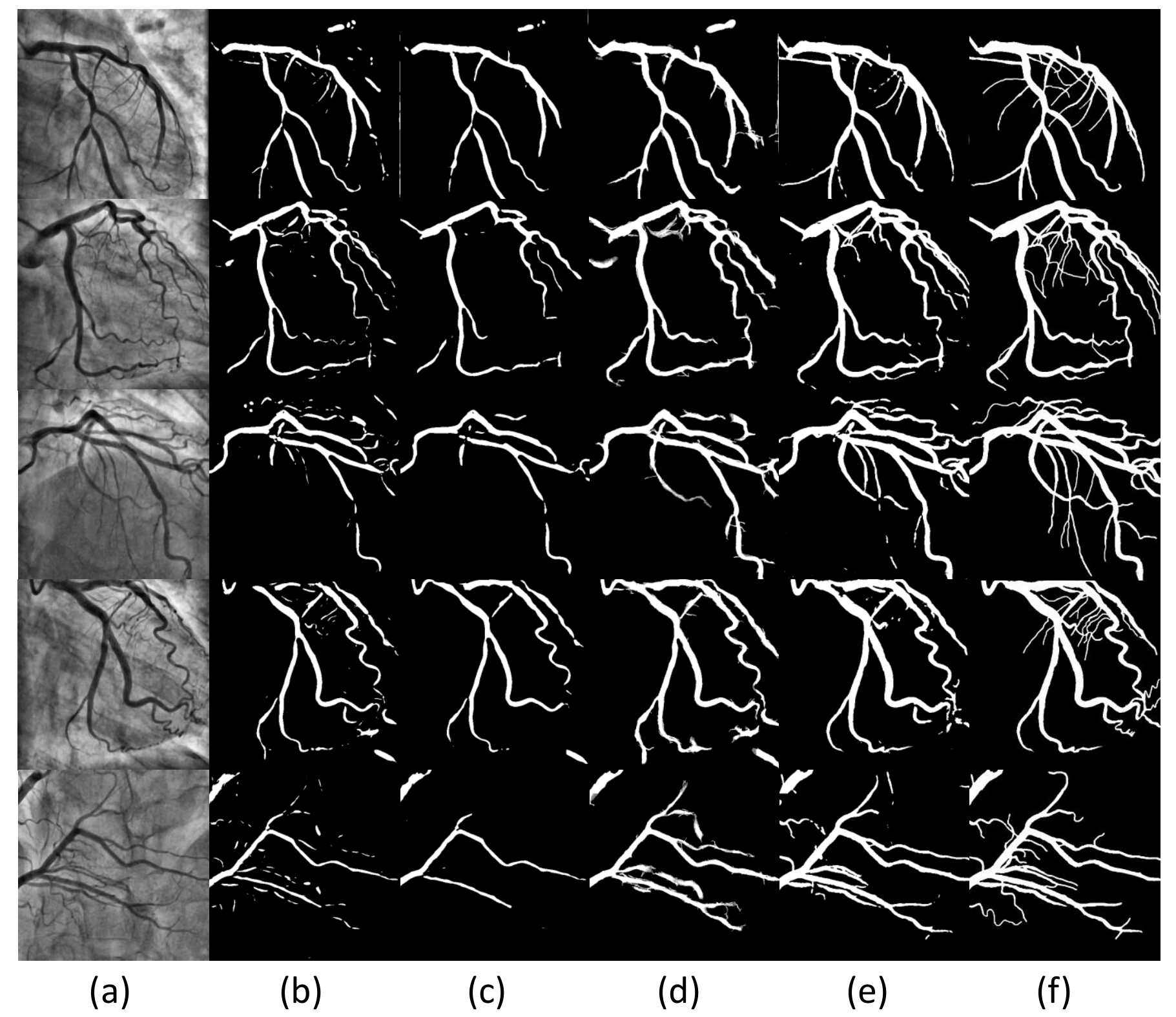}
	\caption{Examples of different vessel segmentation methods. (a) Original images, (b) Frangi Algorithm, (c) Classic U-Net, (d) Add U-Net, (e) Proposed SC-GAN, (f) Ground truth.}
	\label{fig:segmentation}
\end{figure}

In this section, we briefly report the evaluation results in two aspects: 1) Images synthesis and 2) Images segmentation.
\paragraph{Images synthesis:}
Figure \ref{fig:translation} shows some examples of the fundus, DSA, and synthetic image patches. Compared to the results of Cycle-GAN \cite{zhu2017unpaired}, the synthetic images from our proposed SC-GAN have more realistic DSA background and also preserve the vascular structures corresponding to the labels (see the columns (c) and (d) in Figure \ref{fig:translation}).
\paragraph{Images segmentation:}
We annotate 30\% of the DSA dataset (328 out of 1092 images) and evaluate our proposed model on it. Table \ref{fig:segmentation} compares the performance of different methods. As the baseline method of this article, the Frangi algorithm has a Dice score of 0.636$\pm$0.046. If the result of the Frangi algorithm is used as an annotation to train a U-Net (Classic U-Net), the Dice score reduces to 0.589$\pm$0.049. Both Add U-Net and SC-GAN have higher Dice scores (0.742$\pm$0.048 and 0.824$\pm$0.026). And SC-GAN also outperforms the other methods in terms of accuracy and recall. Figure \ref{fig:segmentation} shows some typical examples in the test set. Columns (d-e) show better results than columns (b-c), indicating that knowledge transfer effectively enhances the identification of small blood vessels. By comparing the results of Add U-Net and SC-GAN, we can also find that GAN is better than an average in terms of the quality of knowledge transfer.
\begin{table}
	\begin{center}
		\caption{Quantitative performance of different vessel segmentation methods.}
		\begin{tabular}{ccccc}
			\hline
			Methods & Frangi Algorithm  & Classic U-Net & Add U-Net & \textbf{Proposed} \\
			\hline
			Accuracy  & 0.927$\pm$0.014 & 0.921$\pm$0.015  &  0.940$\pm$0.012 & \textbf{0.953$\pm$0.009}   \\
			Precision  & 0.943$\pm$0.057 & \textbf{0.975$\pm$0.052} & 0.864$\pm$0.078 & 0.820$\pm$0.031  \\
			Recall  & 0.481$\pm$0.048 & 0.423$\pm$0.047 & 0.653$\pm$0.050 & \textbf{0.829$\pm$0.039}  \\
			Dice Coefficient  & 0.636$\pm$0.046 & 0.589$\pm$0.049 & 0.742$\pm$0.048 & \textbf{0.824$\pm$0.026}   \\
			\hline
		\end{tabular}
		\label{tab:segmentation}
	\end{center}
\end{table}

\section{Discussion}
In this paper, we proposed a shape-consistent GAN model (SC-GAN) for coronary artery segmentation, which was able to transfer the knowledge of the segmentation on public fundus dataset to an unlabeled DSA dataset. Experimental results demonstrated that SC-GAN obtained an obvious superior performance on coronary arteries segmentation. Despite the promising results, our method has several limitations and requires further investigation: 1. How well does the method perform on other datasets; 2. Although the segmentor is light-weighted in test, training SC-GAN is much more complex than in a classic supervised deep model. In future work, we will further test the proposed SC-GAN on other application scenarios and simplify the  training process.

\subsubsection{Acknowledgments.} This work was supported by National Key R\&D Program of China (No. 2018YFC0910700); The National Natural Science Foundation of China (NSFC) under Grants 81801778, 11831002; Beijing Natural Science Foundation (Z180001).

%
%
%

\begin{thebibliography}{10}
	\providecommand{\url}[1]{\texttt{#1}}
	\providecommand{\urlprefix}{URL }
	\providecommand{\doi}[1]{https://doi.org/#1}
	
	\bibitem{adams1994seeded}
	Adams, R., Bischof, L.: Seeded region growing. IEEE Transactions on pattern
	analysis and machine intelligence  \textbf{16}(6),  641--647 (1994)
	
	\bibitem{frangi1998multiscale}
	Frangi, A.F., Niessen, W.J., Vincken, K.L., Viergever, M.A.: Multiscale vessel
	enhancement filtering. In: International conference on medical image
	computing and computer-assisted intervention. pp. 130--137. Springer (1998)
	
	\bibitem{ganin2016domain}
	Ganin, Y., Ustinova, E., Ajakan, H., Germain, P., Larochelle, H., Laviolette,
	F., Marchand, M., Lempitsky, V.: Domain-adversarial training of neural
	networks. The Journal of Machine Learning Research  \textbf{17}(1),
	2096--2030 (2016)
	
	\bibitem{hoover2003locating}
	Hoover, A., Goldbaum, M.: Locating the optic nerve in a retinal image using the
	fuzzy convergence of the blood vessels. IEEE transactions on medical imaging
	\textbf{22}(8),  951--958 (2003)
	
	\bibitem{hu2013automated}
	Hu, Q., Abr{\`a}moff, M.D., Garvin, M.K.: Automated separation of binary
	overlapping trees in low-contrast color retinal images. In: International
	conference on medical image computing and computer-assisted intervention. pp.
	436--443. Springer (2013)
	
	\bibitem{ioffe2015batch}
	Ioffe, S., Szegedy, C.: Batch normalization: Accelerating deep network training
	by reducing internal covariate shift. In: International Conference on Machine
	Learning. pp. 448--456 (2015)
	
	\bibitem{kamnitsas2017unsupervised}
	Kamnitsas, K., Baumgartner, C., Ledig, C., Newcombe, V., Simpson, J., Kane, A.,
	Menon, D., Nori, A., Criminisi, A., Rueckert, D., et~al.: Unsupervised domain
	adaptation in brain lesion segmentation with adversarial networks. In:
	International conference on information processing in medical imaging. pp.
	597--609. Springer (2017)
	
	\bibitem{kingma2014adam}
	Kingma, D.P., Ba, J.: Adam: A method for stochastic optimization. arXiv
	preprint arXiv:1412.6980  (2014)
	
	\bibitem{lapin2018analysis}
	Lapin, M., Hein, M., Schiele, B.: Analysis and optimization of loss functions
	for multiclass, top-k, and multilabel classification. IEEE transactions on
	pattern analysis and machine intelligence  \textbf{40}(7),  1533--1554 (2018)
	
	\bibitem{nie2017medical}
	Nie, D., Trullo, R., Lian, J., Petitjean, C., Ruan, S., Wang, Q., Shen, D.:
	Medical image synthesis with context-aware generative adversarial networks.
	In: International Conference on Medical Image Computing and Computer-Assisted
	Intervention. pp. 417--425. Springer (2017)
	
	\bibitem{osher1988fronts}
	Osher, S., Sethian, J.A.: Fronts propagating with curvature-dependent speed:
	algorithms based on hamilton-jacobi formulations. Journal of computational
	physics  \textbf{79}(1),  12--49 (1988)
	
	\bibitem{staal2004ridge}
	Staal, J., Abr{\`a}moff, M.D., Niemeijer, M., Viergever, M.A., Van~Ginneken,
	B.: Ridge-based vessel segmentation in color images of the retina. IEEE
	transactions on medical imaging  \textbf{23}(4),  501--509 (2004)
	
	\bibitem{tsai2018learning}
	Tsai, Y.H., Hung, W.C., Schulter, S., Sohn, K., Yang, M.H., Chandraker, M.:
	Learning to adapt structured output space for semantic segmentation. In:
	Proceedings of the IEEE Conference on Computer Vision and Pattern
	Recognition. pp. 7472--7481 (2018)
	
	\bibitem{ulyanov2016instance}
	Ulyanov, D., Vedaldi, A., Lempitsky, V.: Instance normalization: The missing
	ingredient for fast stylization. arXiv preprint arXiv:1607.08022  (2016)
	
	\bibitem{zhang2017mixup}
	Zhang, H., Cisse, M., Dauphin, Y.N., Lopez-Paz, D.: mixup: Beyond empirical
	risk minimization. arXiv preprint arXiv:1710.09412  (2017)
	
	\bibitem{zhang2018translating}
	Zhang, Z., Yang, L., Zheng, Y.: Translating and segmenting multimodal medical
	volumes with cycle-and shape-consistency generative adversarial network. In:
	Proceedings of the IEEE Conference on Computer Vision and Pattern
	Recognition. pp. 9242--9251 (2018)
	
	\bibitem{zhu2017unpaired}
	Zhu, J.Y., Park, T., Isola, P., Efros, A.A.: Unpaired image-to-image
	translation using cycle-consistent adversarial networks. In: Proceedings of
	the IEEE International Conference on Computer Vision. pp. 2223--2232 (2017)
	
	\bibitem{zuiderveld1994contrast}
	Zuiderveld, K.: Contrast limited adaptive histogram equalization. In: Graphics
	gems IV. pp. 474--485. Academic Press Professional, Inc. (1994)
	
\end{thebibliography}
%
\end{document}